\documentclass[aps,twocolumn,showpacs,prl]{revtex4-1}

\usepackage{hyperref}
\hypersetup{colorlinks=true,linkcolor={blue},citecolor={blue}, urlcolor=blue}

\usepackage{epsfig}
\usepackage{newlfont}
\usepackage{amssymb}
\usepackage{amsfonts}
\usepackage{amsmath}
\usepackage{bm}
\usepackage{cancel}

\def\lsim{\mathrel{\rlap{\lower4pt\hbox{\hskip1pt$\sim$}}
    \raise1pt\hbox{$<$}}}                
\def\gsim{\mathrel{\rlap{\lower4pt\hbox{\hskip1pt$\sim$}}
    \raise1pt\hbox{$>$}}}                

\begin{document}

\renewcommand*{\DefineNamedColor}[4]{%
   \textcolor[named]{#2}{\rule{7mm}{7mm}}\quad
  \texttt{#2}\strut\\}

\definecolor{red}{rgb}{1,0,0}
\definecolor{cyan}{cmyk}{1,0,0,0}

\title{Variational Renormalization Group for Dissipative Spin-Cavity Systems:\\ Periodic Pulses of Nonclassical Photons from Mesoscopic Spin Ensembles}
\author{Himadri Shekhar Dhar, Matthias Zens, Dmitry O. Krimer, and Stefan Rotter}

\affiliation{Institute for Theoretical Physics, Vienna University of Technology (TU Wien), Wiedner Hauptstra{\ss}e 8-10/136, 1040, Vienna, Austria, EU}


\begin{abstract}
Mesoscopic spin ensembles coupled to a cavity offer the exciting prospect of observing complex nonclassical phenomena 
that pool the microscopic features from a few spins with those of macroscopic spin ensembles.
Here, we demonstrate how the collective interactions in an ensemble of as many as hundred spins can be harnessed to obtain a periodic pulse train of nonclassical light. 
To unravel the full quantum dynamics and photon statistics, we develop a time-adaptive variational renormalization group method that accurately captures the underlying Lindbladian dynamics of the mesoscopic spin-cavity system. 
\end{abstract}

\maketitle

\emph{Introduction.--} In the past decade, there has been considerable interest in the development of hybrid quantum systems \cite{Xiang2013,Kurizki2015}, 
where the interactions between 
spins or emitters with the modes of an electromagnetic field offer a cumulative advantage in designing quantum protocols, ranging from  quantum many-body  simulations \cite{Houck2012,Schmidt2013} to the processing and storage of quantum information \cite{Imamoglu1999,Simon2010,Wesenberg2009,Sangouard2011}.
The majority of theoretical and experimental studies on such hybrid systems have focused on two very distinct regimes:
On the one hand, macroscopic spin ensembles (SEs) and their collective properties  have been investigated in the context of superradiance \cite{Dicke1954,Rose2017},
amplitude bistability \cite{Lugiato1984,Angerer2017}, spectral engineering \cite{Riedmatten2008,Krimer2015}, quantum memories \cite{Moiseev2001,Julsgaard2004,Greze2014}, and {suppression of decoherence through the cavity protection effect \cite{Diniz2011,Krimer2014, Zhong2017}.}
%
In this macroscopic limit, however,  the light-matter interaction can be treated already on a semiclassical level \cite{Bonifacio1982}, with possible quantum corrections \cite{Keeling2009,Kramer2015}. 
On the other hand, in the microscopic limit, where a single or just a few spins couple to a cavity, this interaction demands 
full quantum solutions due to the anharmonicities of the excitations \cite{Fink2008}, resulting in exotic nonclassical phenomena such as antibunching \cite{Paul1982}, photon-blockade \cite{Imamoglu1997}, and single-photon emission \cite{Aharonovich2016}.
Here we will explore the largely uncharted mesoscopic regime 
that offers the unique possibility to synergistically combine collective with non-classical features that are otherwise restricted to the two separate regimes mentioned above.  
First signatures in this direction are already starting to emerge, such as through the observation of
{unconventional photon blockade \cite{Flayac2017,Radulaski2017,Blazquez2018,Snijders2018,Vaneph2018}}, superbunching \cite{Jahnke2016} and nonclassical photon bundles \cite{Munoz2014}.

While experimental implementations of mesoscopic SEs are already within reach, especially using superconducting qubits \cite{Macha2014}, quantum dots 
\cite{Strauf2006}, NV centers \cite{Bradac2017}, {rare earth ensembles \cite{Zhong2017b}}, and atomic gases \cite{Rempe1991,Sauer2004},
theoretical studies for such systems have been restricted to very specific regimes, as the exponential growth of the Hilbert space limits any complete solution beyond a few spins. 
Most commonly, one is limited to either very weak excitations \cite{Carmichael1991, Radulaski2017b, Blazquez2017}, few spin systems \cite{Temnov2005,Auffeves2011}, or 
to ensembles without any inhomogeneous broadening \cite{Henschel2010,Leymann2013,Gegg2016, Kirton2017,Armen2006, Mabuchi2008}. Although these limits have already provided valuable insights into mesoscopic systems, they represent only the tip of the iceberg. There is definitely more to explore when going beyond these restrictions by taking into account the  complex interplay between quantum effects, 
inhomogeneity and nonlinearity due to excitations.

In this Letter, we formulate a powerful approach to investigate the full quantum dynamics of an inhomogeneous mesoscopic ensemble of as many as hundred spins inside a quantum cavity, driven by a short coherent field.
The spins are arranged such that their transition frequencies form a spectral frequency comb \cite{Afzelius2009,Zhang2015,Krimer2016}.
We demonstrate that the temporal evolution of such a comb-shaped ensemble results in a periodic and long-lived pulse train of nonclassical photons in the cavity.
Here, the mesoscopic limit allows us to profit from an enhanced collective spin-cavity coupling, while also creating sub-Poissonian light fields due to the anharmonic nature of the excitations.
{In particular, the synergy of anharmonic and collective properties gives rise to periodic photon pulses operating close to the single-photon regime, which provides a valuable resource for quantum protocols such as linear optical quantum computing \cite{Knill2001}, single-photon cryptography \cite{Beveratos2002} and low-light imaging \cite{Morris2015}. In intervals between two photon pulses, the field also exhibits the exotic phenomenon of superbunching, which is often associated with correlations in the spin ensemble \cite{Jahnke2016} or in the gain medium of quantum-dot microlasers \cite{Leymann2013b}.  
Moreover, the strong driving in this regime also provides the exciting prospect of creating relatively high cavity photon numbers with nonclassical statistics. 
%
In turn, however, the corresponding spin-cavity dynamics cannot be analyzed using known theoretical approaches. 
We thus develop a time-adaptive variational renormalization group method \cite{Verstraete2008,Schollwoeck2011,Orus2014} that efficiently describes the Lindbladian dynamics of the mesoscopic spin-cavity system. For a coherent reading, we begin with a description of our model and the resulting physics before presenting our method.
}

%

\begin{figure}[t]
\includegraphics[width=3.3in,angle=00]{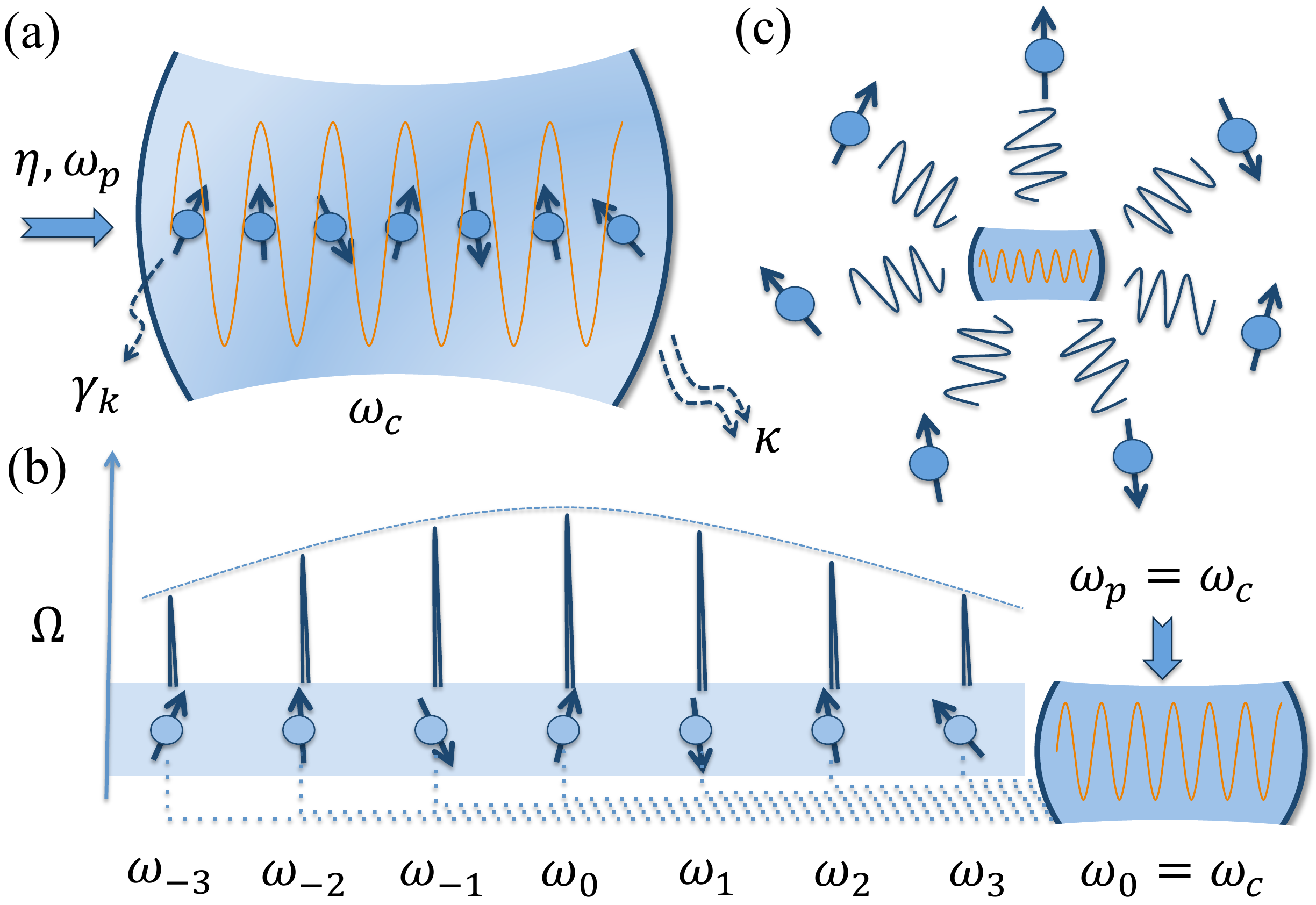}
\caption{(a) A mesoscopic spin ensemble interacting with a cavity, resonant at frequency $\omega_c$. The cavity is driven by a coherent pulse of strength $\eta$ and frequency $\omega_p$. The cavity and spin losses are given by $\kappa$ and $\gamma_k$. {(b) The spin transition frequencies $\omega_j$ in the ensemble form a spectral frequency comb, with collective coupling strength $\Omega$. (c) The spin-cavity system can be considered as a central body system.}}
\label{fig1}
\end{figure}

\emph{Model.--}
An ensemble of $N$ two-level emitters or spins inside a cavity, see Fig.~\ref{fig1}(a), 
can be modelled using the Tavis-Cummings Hamiltonian \cite{Tavis1968}, which
under the dipole and rotating wave approximations reads,

\begin{eqnarray}
\mathcal{H} &=& \frac{1}{2} \sum_{k=1}^{N} \omega'_{k} ~ \sigma^z_k  + \omega_{c}~ \hat{a}^\dag_c \hat{a}_c +i\sum_{k=1}^{N} g_k(\sigma^+_k \hat{a}_c - \sigma^-_k \hat{a}^\dag_c)\nonumber\\
&&+ ~ i (\eta(t)~\hat{a}^\dag_c e^{-i\omega_p t} - \eta^*(t)~\hat{a}_c e^{i\omega_p t})\,,
\label{Ham}
\end{eqnarray}
{where we take $\hbar = 1$. Here $\omega_c$ is} the resonance frequency of the cavity field $\hat{a}_c$, and $\omega'_{k}$, $g_k$ are the transition frequency and coupling strength for the $k^{th}$ spin (a spatial dependence of the spins can be included, but is not considered here). Furthermore, $\sigma^z_k$, $\sigma^+_k$ and $\sigma^-_k$ are the spin-{\footnotesize 1/2} Pauli operators. The quantum cavity is coherently driven with frequency $\omega_p$ and intensity $\eta(t)$.
In general the cavity and the spins in the mesoscopic ensemble are lossy, and the open dynamics is  governed by the Lindblad equation,  
$
{d\rho}/{dt} =\mathcal{L}[\rho] = -i[\mathcal{H},\rho]+ \kappa \mathcal{L}_{\hat{a}_c}[\rho]+ \sum_k \gamma_k \mathcal{L}_{\sigma^-_k}[\rho],
$
where $\mathcal{L}_{\hat{x}}[\rho] =  \hat{x} \rho {\hat{x}}^\dag -\frac{1}{2} \{ \hat{x}^\dag \hat{x}, \rho\}$ for $\hat{x} = \hat{a}_c$ and $\sigma^-_k$.
The radiative losses of the cavity and spins are given by $\kappa$ and $\gamma_k$. 
{For macroscopic SEs, the spins and the cavity can both be treated semiclassically, and the expectation values are solved using the Maxwell-Bloch equations \cite{Bonifacio1982}, and quantum corrections thereof \cite{Keeling2009,Kramer2015}. However, to capture all the complex features of the quantum dynamics in mesoscopic systems, the Lindblad equation needs to be solved exactly. }

{\emph{Nonclassical light in mesoscopic ensembles.--} To demonstrate the complex nonclassical  phenomena associated with mesoscopic spin-cavity interactions, we consider ensembles with up to $N$ = 105 spins arranged in a finite spectral comb, with transition frequencies spaced at equidistant intervals.} Such frequency combs have already been engineered in macroscopic ensembles, where the collective interactions result in long coherence times suitable for efficient quantum memory protocols \cite{Afzelius2009,Zhang2015} and long-lived pulses of classical light \cite{Krimer2016}. 
We demonstrate now explicitly that moving to a mesoscopic SE allows us to harness the quantum effects of light-matter interactions thus making the pulses emitted from the mesoscopic SE nonclassical. Specifically, we propose a protocol for creating
a periodic pulse train of antibunched light with sub-Poissonian photon statistics.
\begin{figure}[b]
\includegraphics[width=3.4in,angle=00]{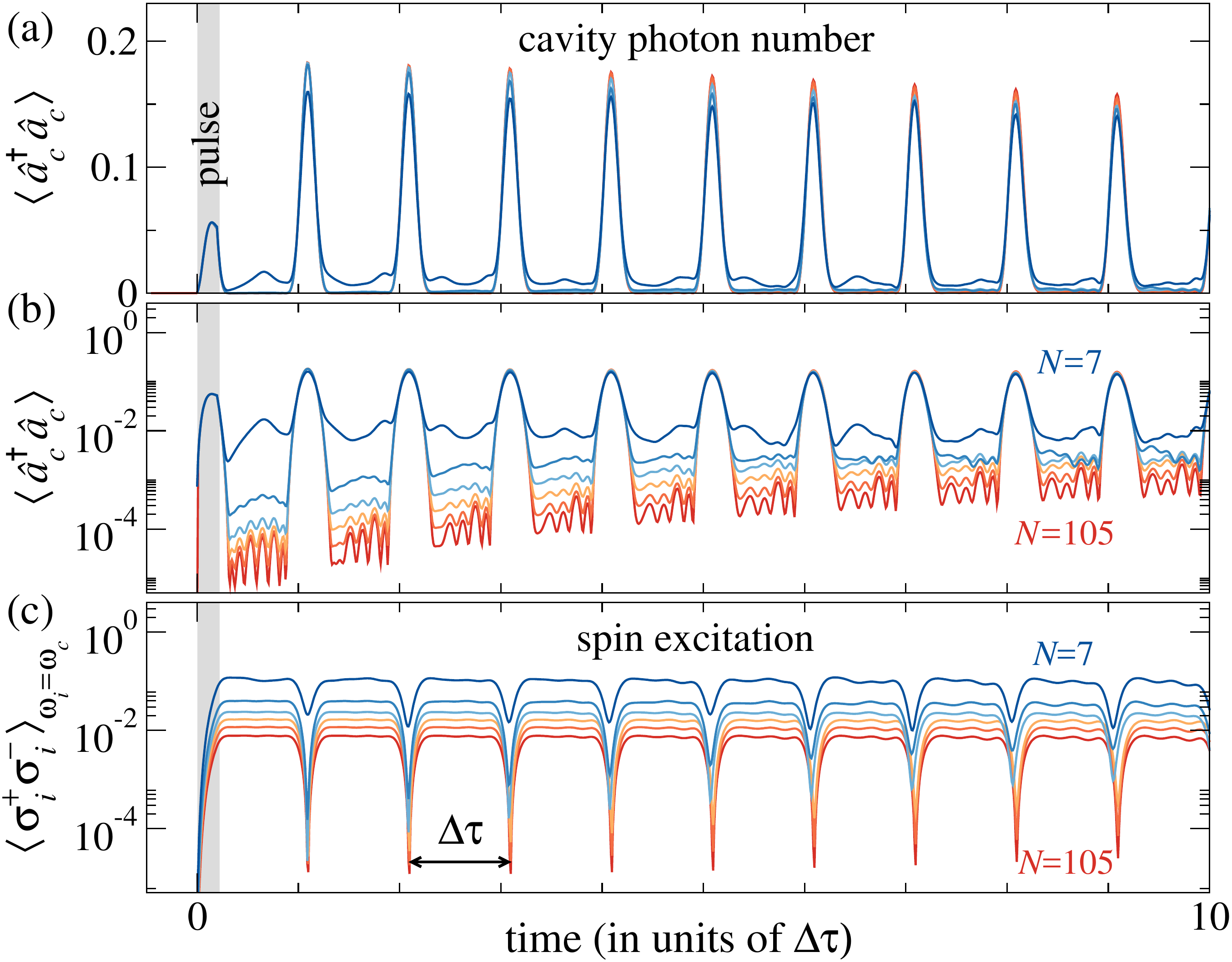}
\caption{Temporal evolution of the mesoscopic spin-cavity system. The figures show  the cavity photon number $\langle\hat{a}_c^\dag\hat{a}_c\rangle$ in (a) linear and (b) log scale, and (c) the spin excitation at resonance $\langle \sigma_i^+ \sigma_i^-\rangle_{\omega_i=\omega_c}$, varying with time, and for ensembles containing $N$ = 7, 21, 35, 49, 70, and 105 spins, shown with colors varying from blue to red. The shaded region at times $0\leq t \leq t'$ indicates the short rectangular driving pulse and $\Delta\tau$ is the interval between the periodic revivals.}
\label{fig2}
\end{figure}
The 
transition frequencies ($\omega'_{k}$) of the spins in the spectral comb are arranged around the cavity frequency, $\omega_c$, with $m$ (odd) distinct frequencies given by, 
$\omega_j$ = $\omega_c + j \Delta\omega$, for $j = \{-(m-1)/2,\dots, (m-1)/2\}\}$, as shown in Fig.~\ref{fig1}(b). 
For an $N$-spin ensemble inside the cavity, each frequency $\omega_j$ in the spectral comb corresponds to a subensemble of $N'$ = $N/m$ spins.
The coupling constants for each of the subensembles and the cavity 
follow a Gaussian distribution,
$\Omega_j = \Omega_0\exp\left[-(\omega_c-\omega_j)^2/2\lambda^2\right]$, where $\Omega_0$ is the coupling strength for the central subensemble, which is resonant with the cavity, and $\lambda$ is the standard deviation of the distribution.
Assuming that within each of the altogether $m$ = 7 subensembles all the spins have the same coupling strength, i.e., $\Omega_j^2$ = $\sum_k^{N'} g_{j,k}^2 = N'g_j^2$, the collective coupling of the total spin ensemble is given by $\Omega^2$ = $\sum_k \Omega_k^2$. We drive this hybrid quantum system resonantly with a short coherent pulse of intensity, $\eta(t) = \eta \in \Re$, for $0\leq t\leq t'$, and $\eta(t) = 0$, otherwise. 
%
Before the pulse arrives, the initial spin-cavity system is unexcited and the cavity is in the vacuum state. 
The 
coupling strengths and characteristic width of the comb 
are chosen as
$\Omega_0/2\pi$ = 30 MHz, $\lambda/2\pi$ = 150 MHz, and $\Delta\omega/2\pi$ = 40 MHz. 
The cavity and spin loss terms are taken as $\kappa/2\pi$ = 0.4 MHz, and $\gamma$ = $\kappa/40$, respectively, with
$\eta$ = 40$\kappa$. {The driving pulse duration $t'$ is $1/5$ of the characteristic timescale $2\pi/\Delta \omega$.}


{The first important feature of our mesoscopic frequency comb is 
the periodic pulse train of light it emits, exhibited by sharp revivals of the average cavity photon number, $\langle \hat{a}^\dag_c\hat{a}_c\rangle$, as shown in Fig.~\ref{fig2}.}
Here, the peaks correspond to the collective transfer of excitations from the spin ensemble to the cavity, as evident from the sharp decrease in the spin excitation $\langle \sigma^+_i\sigma^-_i\rangle_{\omega_i = \omega_c}$ at the revivals. {The periodic pulses result from the constructive rephasing of spins in different subensembles of the comb, with the time interval
between subsequent peaks commensurate with the inverse of the spectral width, $\Delta\tau \approx 2.2\pi/\Delta\omega = 173$ ns. This is a hallmark of the collective behavior of the spins in the spectral frequency comb \cite{comment}.}
We note that during the transfer of energy from the cavity to the ensemble, {larger ensembles, i.e., larger $N$, not only lead to enhanced coupling but also produce more stable and sharper photon pulses as excitations are distributed over more spins. In contrast,} for few spins, significant photon excitations may also exist between the peaks, as observed in Fig.~\ref{fig2}(b).
%
%

\begin{figure}[t]
\includegraphics[width=3.4in,angle=00]{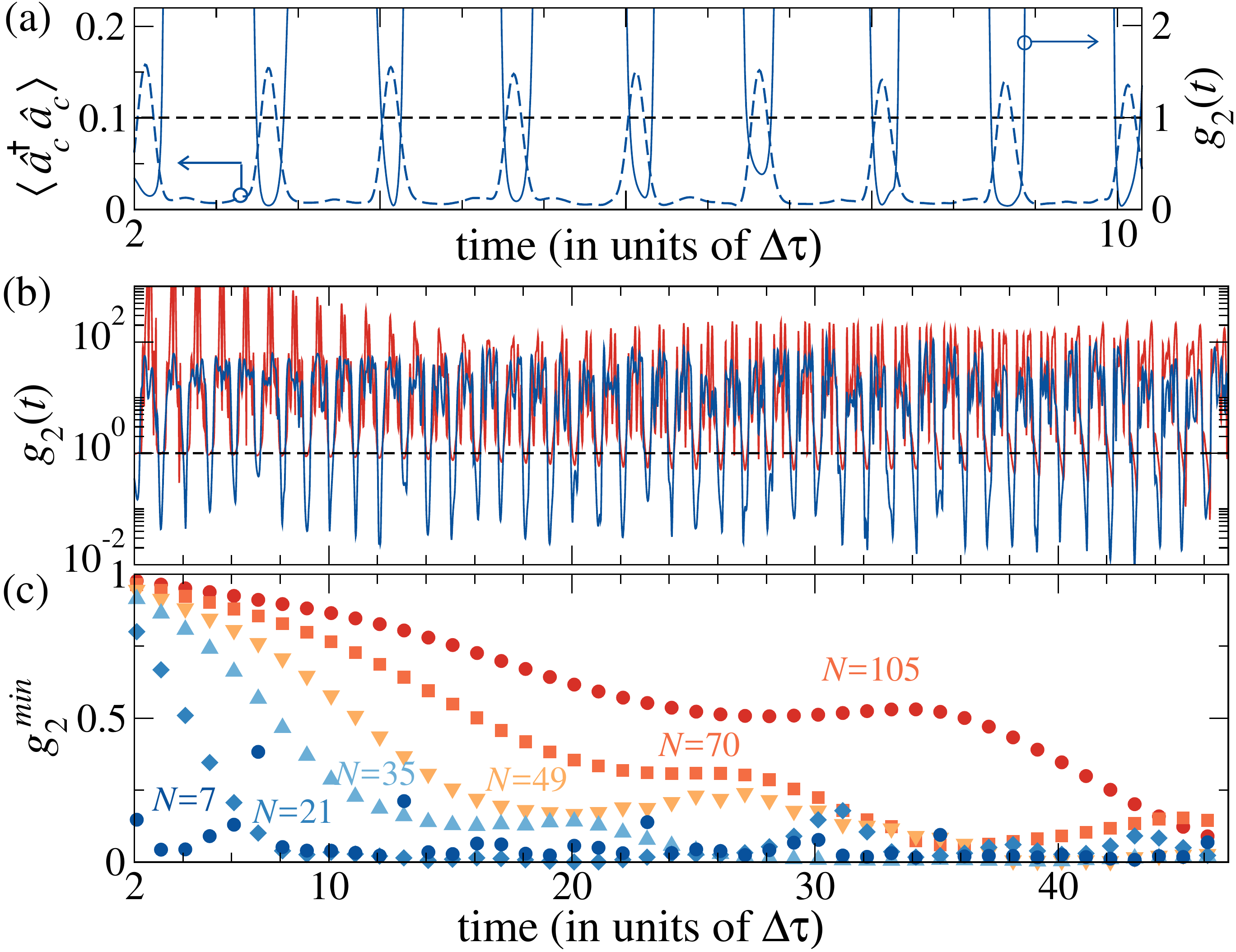}
\caption{Photon statistics of the transient cavity field (all parameters and color codings are the same as in Fig.~\ref{fig2}). (a) Equal-time second-order correlation function $g_2(t)$ in comparison with the cavity photon number $\langle\hat{a}^\dag_c\hat{a}_c\rangle$, for  $N$ = 7. (b) Temporal evolution of  $g_2(t)$ for $N$ = 7 and 105. (c) Minimum value of $g_2(t)$ close to a pulse revival, $g_2^{min}$, for $N$ = 7, 21, 35, 49, 70, and 105. 
The horizontal black-dashed line in (a) and (b) corresponds to $g_2(t)$ = 1 for coherent light.}
\label{fig3}
\end{figure}

{The second 
important, and in fact, central feature of these periodic light pulses is their distinct nonclassical character as inherent in their photon statistics. Using the equal-time second order correlation function at time $t$, defined as 
$g_2(t)$ = $\langle \hat{a}^{\dag 2}_c(t) \hat{a}_c^2(t)\rangle$/$\langle \hat{a}^\dag_c(t)\hat{a}_c(t)\rangle^2$,
we observe in Fig.~\ref{fig3} that at times where the photon pulse arrives the field is distinctly sub-Poissonian, i.e.,  $g_2(t)<1$.}
While for all classical sources $g_2(t) \geq$ 1 (unity for coherent light), sub-Poissonian light with $g_2(t) <$ 1 is explicitly nonclassical. 
Here, we observe that such nonclassicality of the pulse train persists even for ensembles containing more than hundred spins, as evidenced by $g_2^{min}<1$ in Fig.~\ref{fig3}(c).
Therefore, a relatively large mesoscopic ensemble can be used to generate a high quality, nonclassical pulse train of photons.
We note that there are interesting demarcations in the nonclassical nature of the photon pulse. To be specific, while the pulse are sub-Poissonian and distinctly nonclassical at all times (for all $N$), the unambiguous single-photon regime, $g_2^{min}\ll 1$, is achieved only after a finite evolution time (which is shorter for smaller $N$). 
%

We also explicitly checked
that higher-order correlation functions, up to order $n=4$, are also less than unity, i.e., $g_n(t)$ = $\langle \hat{a}^{\dag n}_c(t) \hat{a}_c^n(t)\rangle$/$\langle \hat{a}^{\dag}_c(t)\hat{a}_c(t)\rangle^n<1$. {Interestingly, at $t > 45\Delta\tau$, very low values of $g_n(t)$ ($n$ = $2,3,4$) are attained even for $N = 105$ (see Fig.~1 of the supplemental material \cite{supp}).} 
%
%
In an experimental setting, such a suppression 
of multi-photon detection 
is considered to be a distinctive {characteristic} of single-photon emitters \cite{Carreno2016}, {making our system an interesting candidate for the design of quantum protocols \cite{Knill2001,Beveratos2002,Morris2015}.}
An important feature is the persistent periodicity $\Delta\tau$ of the nonclassical pulse, which can be modulated by tuning the peak spacing $\Delta\omega$ in the spectral frequency comb. {We note that this periodicity is present even for larger ensembles where antibunching is weak.} 
For hybrid quantum systems, such a periodicity may allow for a temporal synchronization of the nonclassical light during an experimental phase, which is crucial for quantum memory protocols \cite{Julsgaard2004}. 
%
Another interesting cooperative behavior we observe is superbunching of the cavity field  in intervals between the peaks, where $g_2(t)\gg 1$.
This phenomenon has previously been related to superradiance arising from spin correlations in the ensemble \cite{Jahnke2016, Bhatti2015}. Here, the low photon number in the superbunched emission is characteristic of the superradiant excitation being collectively stored in the spins rather than in the cavity. {Such effects have also been reported in bimodal quantum-dot microlasers, where superbunching is induced by correlations in the gain medium \cite{Leymann2013b} or irregular mode-switching \cite{Redlich2016}.}

{\emph{Renormalization for the Lindbladian dynamics.--} 
To arrive at the above results we take advantage of the fact that several exotic phenomena in spin-ensemble-cavity systems arise from Lindbladian dynamics that does not
necessarily generate large correlations between the spins. This allows us to apply a time-adaptive variational renormalization group method \cite{Verstraete2008,Schollwoeck2011,Orus2014}, which efficiently maps the transient dynamics to a highly reduced vector space \cite{Zwolak2004, Verstraete2004, Cui2015,Mascarenhas2015}.
%
To set up this approach, we first map the system to a higher-dimensional complex vector space, such that  a $d\times d$ density matrix, $\rho$, is vectorized to a $d^2$ \emph{superket}, $|\rho\rangle=\textbf{vec}(\rho)$. Also, a $d\times d$ operator, $\hat{O}$, which acts on $\rho$, is given by the higher-dimensional, $d^2\times d^2$ \textit{superoperator}, $\hat{\mathcal{O}}$, which now acts on $|\rho\rangle$:  
$\rho \rightarrow |\rho\rangle$,
$\hat{O}\rho \rightarrow (\hat{O}\otimes \mathbb{I})|\rho\rangle = \hat{\mathcal{O}}|\rho\rangle$, and $\rho  \hat{O} \rightarrow (\mathbb{I}\otimes \hat{O}^T )|\rho\rangle = \hat{\mathcal{O}}'|\rho\rangle$.  
The Lindblad equation in such a superoperator space is then given by ${d|\rho\rangle}/{dt} =\tilde{\mathcal{L}}|\rho\rangle$, where
\begin{equation}
\tilde{\mathcal{L}} = -i(\mathcal{H} \otimes \mathbb{I} - \mathbb{I} \otimes \mathcal{H}^T) +  \kappa  \tilde{\mathcal{L}'}_{\hat{a}_c}+  \sum\mathop{}_{\mkern-5mu k} \gamma_k  \tilde{\mathcal{L}'}_{\sigma^-_k},
\label{Lind}
\end{equation}
with $\tilde{\mathcal{L}'}_{\hat{x}} =  \hat{x} \otimes {\hat{x}}^* - \frac{1}{2}\hat{x}^\dag \hat{x} \otimes \mathbb{I} - \frac{1}{2}\mathbb{I} \otimes \hat{x}^T\hat{x}^*$. 
We note that in contrast to low-dimensional quantum spin systems with short-range interactions, all spins in the mesoscopic SE interact only with the cavity. 
Therefore we can map the  renormalization of the hybrid spin-cavity system to a central body problem \cite{Stanek2013}, as illustrated in Fig.~\ref{fig1}(c), albeit in a higher-dimensional superoperator space.
Here, the cavity acts as the central quantum object
that couples to each of the spins in the ensemble, which behaves like a bath in the superoperator space. The terms in $\mathcal{H}$ and $\tilde{\mathcal{L}'}_{\hat{x}}$ [see Eq.~(\ref{Lind})] can thus be written as a sum of individual spin-cavity terms, such that $\tilde{\mathcal{L}} = \sum_k \tilde{\mathcal{L}}_k$. A description of the mapping to the superoperator space is provided in Sec. IA of the supplemental material \cite{supp}.}

The two key steps in implementing the variational renormalization group for the Linbladian dynamics are (i) the variational search for a truncated superoperator space, and (ii) a time-adaptive Lindbladian evolution of the spin-cavity system. The former is obtained using the Schmidt decomposition,
$|\rho\rangle=\sum_{\tilde{k}=1}^\mathcal{K} \alpha_{\tilde{k}} |\tilde{k}_A\rangle |\tilde{k}_B\rangle$, where the system is divided into blocks, $A$ and $B$, as done during a renormalization method \cite{Schollwoeck2011}. Here, $\{\alpha_{\tilde{k}}\}$ are the Schmidt coefficients in descending order, and $|\tilde{k}_A\rangle$ and $|\tilde{k}_B\rangle$ are the eigenvectors of the reduced superoperators of $|\rho\rangle$.
$\mathcal{K}$ is bounded from above by $r$ = $\textrm{min}[d_{\mathcal{R}_{A}},d_{\mathcal{R}_{B}}]$, and is a measure of the total bipartite correlations \cite{Datta2007}.
{Importantly, for several open systems $\alpha_{\tilde{k}}$ decays rapidly with $\tilde{k}$ \cite{Zwolak2004}. Thus, by retaining only the $\mathcal{D}$ highest values of $\alpha_{\tilde{k}}$, we can approximate $|\rho\rangle$ and renormalize it to a significantly reduced dimension  i.e.,
$|\tilde{\rho}\rangle = \sum_{\tilde{k}=1}^\mathcal{D} \alpha_{\tilde{k}}~ |\tilde{k}_A\rangle|\tilde{k}_B\rangle$, where $\mathcal{D} \ll r$. The accuracy of the renormalization depends on the choice of $\mathcal{D}$ and is exact for weakly or uncorrelated systems. For very high correlations, large values of $\mathcal{D}$ need to be considered and the method is less efficient.}
{To implement the Lindbladian evolution, we consider the dynamics governed by $d|\rho\rangle/dt = \sum_k \tilde{\mathcal{L}}_k|\rho\rangle$. 
The superoperator space of the system is numerically renormalized and truncated at each step in a time-adpative manner, using the Suzuki-Trotter decomposition \cite{Suzuki1990}.
This approach is comparable to a time-evolving block decimation \cite{Vidal2003,Vidal2004} or a time-dependent density matrix renormalization group \cite{Daley2004,White2004}.}
%
{A detailed description of our method, error analysis and a benchmark against exact solutions for few spins is provided in Secs.~I--IV of the supplemental material \cite{supp}.
}

{For mesoscopic SEs in a cavity, solutions for the quantum dynamics have so far been achieved only for few limited cases such as for very weak excitations, where 
only a couple of low-excitation states are considered \cite{Carmichael1991}. 
Alternatively, $\mathcal{L}[\rho]$ can be approximated by an effective Hamiltonian \cite{Radulaski2017b,Blazquez2017} in the weak excitation regime where quantum jumps are neglected. In turn, quantum trajectories include jumps but are limited to few spins \cite{Temnov2005,Auffeves2011}.}
Other methods involve direct solutions of $\mathcal{L}[\rho]$, using permutation symmetry for ensembles of identical spins \cite{Gegg2016, Kirton2017}, cumulant expansions for weakly correlated homogeneous ensembles \cite{Henschel2010, Leymann2013}, or approximate semiclassical solutions \cite{Armen2006, Mabuchi2008}.
For the renormalization method we develop, the evolution is decomposed and exactly solved at the level of individual spin-cavity terms.
This allows us to work in an extended parameter regime, with far more spins, higher number of excitations and with inhomogeneous ensembles, {which are typically not accessible using one of the above methods.}
Moreover, being based on the seminal Lindblad equation, our approach is distinct from those tensor-network methods that study open dynamics by simulating the unitary evolution of the larger system-environment states \cite{Prior2010,Schroeder2016,Pino2018}. In particular, our approach 
does not require any additional restrictions on the environment beyond the Master equation formalism. Our method is thus a powerful tool to obtain the transient or steady states of mesoscopic spin-cavity systems.

\emph{Conclusion and outlook.--} 
We demonstrate that mesoscopic ensembles of spins coupled to a quantum cavity provide an interesting new platform for studying and tailoring non-classical light fields.
%
Based on recent experimental progress \cite{Xiang2013, Kurizki2015}, 
implementing the proposed comb-shaped ensemble
should be readily possible and an attractive option for creating a pulsed quantum source of light.  
These results provide just a first glimpse into the complex quantum dynamics of mesoscopic spin cavity systems now accessible with the numerical method we introduce here. Our approach is based on the key insight that
variational renormalization group and tensor network methods that have recently been successfully 
applied to low-dimensional quantum many-body systems \cite{Schollwoeck2011,Verstraete2008,Orus2014},
can be adapted to 
efficiently treat open spin-cavity systems. 
We thereby bridge a gap in the theoretical understanding of mesoscopic spin ensembles, and open up new directions to investigate complex parameter regimes that have remained out of reach so far. 

\begin{acknowledgments}
We thank M. Liertzer, F. Mintert,  L.A. Orozco, S.S. Roy, and A. Schumer, for fruitful discussions.  We acknowledge support by the Austrian Science Fund (FWF) through the Lise Meitner programme, Project No. M~2022-N27 and the European Commission under project NHQWAVE (MSCA-RISE 691209). The computational results presented have been achieved using the Vienna Scientific Cluster (VSC).
\end{acknowledgments}

\pagebreak

\begin{widetext}

\section*{\large{SUPPLEMENTAL MATERIAL}\\~~\\}

\section{I. Temporal dynamics of the mesoscopic spin ensemble-cavity system}

Over the past few decades variational renormalization group methods based on a broader tensor-network formalism \cite{Schollwoeck2011,Verstraete2008,Orus2014} have been very successful in describing the ground states and unitary dynamics of one-dimensional  many-body quantum systems, especially in the context of properties and critical phenomena related to strongly-correlated systems. In recent years these methods have been extended to study higher-dimensional  \cite{Verstraete2008,Orus2014} as well as finite temperature mixed states \cite{Zwolak2004,Verstraete2004} in quantum spin lattices.

In the main text, we introduce a time-adaptive variational renormalization group method to study the Lindbladian evolution of hybrid quantum systems, with a mesoscopic number of emitters or spins interacting with the modes of a quantum cavity. Unlike quantum spin models with short-range (or fast-decaying) spin-spin interaction, which are often the subject of conventional tensor-network methods \cite{Schollwoeck2011, Verstraete2008, Orus2014, Zwolak2004, Verstraete2004}, 
direct interactions among spins are typically neglected in spin ensemble-cavity systems.
On the contrary, all spins in the ensemble interact with the quantum cavity, and the hybrid spin-cavity system can be treated like a central spin problem \cite{Stanek2013}, albeit with individual decoherences, such that the total system dynamics is that of an open system. In our approach, we treat the cavity like a central quantum system, written in the Fock basis, which interacts with the spins in the ensemble that act like a fermionic bath. Starting from a Lindbladian that describes the temporal dynamics of the mesoscopic ensemble-cavity system, the reduced superoperator space of the 
spins is renormalized and truncated at each step in a time-adaptive manner, similar to a time-evolving block decimation (TEBD) \cite{Vidal2003,Vidal2004} or a time-dependent density matrix renormalization group (t-DMRG) method \cite{Daley2004,White2004}. The superoperator space of the cavity is stored exactly while the weakly correlated spin-ensemble space is truncated, thus allowing us to capture
the individual spin-cavity interactions more accurately. The novelty of our approach is that the time-adaptive renormalization is done directly in the superoperator space that describes the Lindbladian Master equation, instead of the more widely implemented approach that simulates the unitary time evolution by renormalizing the much larger but restricted system-environment Hilbert space \cite{Prior2010,Schroeder2016,Pino2018}.

\subsection{A. The mesoscopic spin ensemble-cavity system}
In general, spin-cavity interactions can be modelled using the Tavis-Cummings (TC) Hamiltonian, as defined in Eq.~(\ref{Ham}) of the main text. For an $N$-spin ensemble, this can be expressed in terms of individual spin-cavity interactions,
\begin{eqnarray}
\mathcal{H} &=& \underbrace{\frac{1}{2} \sum_{k=1}^{N} \omega_{k} ~ \sigma^z_k}_{\sum_{k=1}^{N}\mathcal{H}_{s,k}}  + \underbrace{\omega_{c}~ \hat{a}^\dag_c \hat{a}_c
- i(\eta(t)~a^\dag e^{-i\omega_p t} - \eta^*(t)~a e^{i\omega_p t})\vphantom{\sum_{k=1}^{N}}}_{\mathcal{H}_c} + \underbrace{i\sum_{k=1}^{N} g_k(\sigma^+_k \hat{a}_c + \sigma^-_k \hat{a}^\dag_c)}_{\sum_{k=1}^{N}\mathcal{H}_{int,k}}.
\label{eq1}
\end{eqnarray}
where, we have considered the same assumptions and set of system parameters to define the Hamiltonian, as in the main text.
%
The TC Hamiltonian can thus be recast as,
\begin{equation}
\mathcal{H} = \sum_{k=1}^{N} \mathcal{H}_k = \sum_{k=1}^{N} (\mathcal{H}_{s,k} + \mathcal{H}_{int,k} + \frac{1}{N}\mathcal{H}_c).
\end{equation}

In general the spins and the cavity in the mesoscopic ensemble are lossy, which arises from spontaneous emission, imperfections in the cavity, and non-radiative losses due to the larger environment. These need to be accounted for in the description of the system. In a Markovian setting, such losses in an open system can be described by using a Lindblad master equation of the form 
\begin{equation}
{d\rho}/{dt} =\mathcal{L}[\rho] = -i[\mathcal{H},\rho]+\frac{1}{2} \sum_l \tilde{\gamma}_l (2 \hat{L}_l \rho {\hat{L}_l}^\dag - \hat{L}_l^\dag \hat{L}_l \rho - \rho \hat{L}_l^\dag \hat{L}_l ),
\label{sm_Lind1}
\end{equation}
where, $\hat{L}_l$ are the different Lindblad operators. In a cavity QED system, to account for losses in the cavity, one phenomenologically sets a decay term $\tilde{\gamma}_1 = \kappa$, with $\hat{L}_1 = \hat{a}_c$ being the corresponding Lindblad operator. Similarly, spontaneous emission from spins can be accounted for by considering (for the $k^{th}$ spin), $\tilde{\gamma}_{k+1} = \gamma_k$ and $\hat{L}_{k+1} = \sigma^-_k, ~\forall~k$. Additional losses due to non-radiative dephasing of the spins can also be incorporated in a similar manner by choosing a suitable $\hat{L}_l$. For instance, $\tilde{\gamma}_{N+k+1} = \gamma^p_k$ and $\hat{L}_{N+k+1} = \sigma^z_k, ~\forall~k$. We set $\gamma^p = 0$, for the present case, without loosing generality. The Linblad equation is then the same as the one expressed in the main text,
\begin{equation}
{d\rho}/{dt} =\mathcal{L}[\rho] = -i[\mathcal{H},\rho]
+\underbrace{\frac{1}{2} \sum_{k=1}^{N} \gamma_k (2 \sigma^-_k \rho \sigma^+_k - \sigma^+_k\sigma^-_k \rho - \rho \sigma^+_k\sigma^-_k )}_{\sum_{k=1}^{N}\mathcal{L}_{s,k}[\rho]}
+\underbrace{\frac{1}{2} \kappa~ (2 \hat{a}_c \rho \hat{a}_c^\dag - \hat{a}_c^\dag\hat{a}_c \rho - \rho \hat{a}_c^\dag\hat{a}_c)\vphantom{\sum_{j=1}^{N}}}_{\mathcal{L}_{c}[\rho]}.
\label{sm_Lind2}
\end{equation}

As discussed in the main text, the implementation of the variational renormalization group in the open dynamics of the mesoscopic spin ensemble cavity is facilitated by the representation of the system in the higher-dimensional superoperator space. In other words, a $d \times d$ density matrix, $\rho$, is represented as a $d^2$ vector or \emph{superket}, $|\rho\rangle$, and a $d \times d$ operator, $\hat{O}$, 
is given by the higher-dimensional, $d^2 \times d^2$ \textit{superoperator}, $\hat{\mathcal{O}}$.
This mapping to the superoperator space is given by,
\begin{eqnarray}
\rho = \sum_{i,j=1}^d p_{i,j} |i\rangle\langle j| ~&~~\rightarrow~~&~ |\rho\rangle=\sum_{i,j=1}^d p_{i,j} |ij\rangle,\label{m1}\\
\hat{O}~\rho~&~~\rightarrow~~&~ (\hat{O}\otimes\mathbb{I}_d)~|\rho\rangle = \hat{\mathcal{O}}~|\rho\rangle,\label{m2}\\
\rho~\hat{O}~&~~\rightarrow~~&~ (\mathbb{I}_d\otimes\hat{O}^T)~|\rho\rangle = \hat{\mathcal{O}}'~|\rho\rangle,\label{m3}\\
\left[\mathcal{H},\rho\right] = \mathcal{H}~\rho - \rho~\mathcal{H} ~&~~\rightarrow~~&~ (\mathcal{H} \otimes \mathbb{I} - \mathbb{I} \otimes \mathcal{H}^T)~|\rho\rangle,~~\textrm{and} \label{m4}\\
2 \hat{L}_l \rho {\hat{L}_l}^\dag - \hat{L}_l^\dag \hat{L}_l \rho - \rho \hat{L}_l^\dag \hat{L}_l ~&~~\rightarrow~~&~ (2 \hat{L}_l \otimes {\hat{L}_l}^* - \hat{L}_l^\dag \hat{L}_l \otimes \mathbb{I} - \mathbb{I} \otimes \hat{L}_l^T\hat{L}_l^*)~|\rho\rangle.\label{m5}
\end{eqnarray}
In this superoperator space, the Linbladian that describes the open dynamics of the system, is given by ${d|\rho\rangle}/{dt} =\tilde{\mathcal{L}}|\rho\rangle$
with (see Eq.~(\ref{Lind}) in the main text)
\begin{eqnarray}
&\tilde{\mathcal{L}} = 
\underbrace{\sum_{k=1}^{N} [-i(\mathcal{H}_k \otimes \mathbb{I} - \mathbb{I} \otimes \mathcal{H}_k^T)+\frac{\gamma_k }{2} (2 \sigma^-_k \otimes \sigma^{-*}_k - \sigma^+_k\sigma^-_k \otimes \mathbb{I} - \mathbb{I} \otimes \sigma^+_k\sigma^-_k )]}_{\sum_{k=1}^{N}\tilde{\mathcal{L}}_{s,k}}
+\underbrace{\frac{\kappa}{2} (2 \hat{a}_c \otimes \hat{a}_c^* - \hat{a}_c^\dag\hat{a}_c \otimes \mathbb{I} - \mathbb{I} \otimes \hat{a}_c^\dag\hat{a}_c)\vphantom{\sum_{j=1}^{N}}}_{\tilde{\mathcal{L}}_{c}},~~\\
&\textrm{which can be written as}~\tilde{\mathcal{L}} = \sum_k \tilde{\mathcal{L}}_k = \sum_k (\tilde{\mathcal{L}}_{s,k}+\frac{1}{N}\tilde{\mathcal{L}_c}).
\label{Lind2}
\end{eqnarray}
The trace of the density matrix can be defined using the trace superket, $|t\rangle$, such that $\textrm{Tr}[\rho]$ = $\langle t|\rho\rangle$ = 1. Here, $|t\rangle$ is the vector form of the identity operator in the relevant dimension. 
In the superoperator space, the expectation value of some operator $\hat{O}$ can be written as $\langle\hat{O}\rangle$ = $\langle t|\hat{O} \otimes \mathbb{I} |\rho\rangle$, where $\hat{O}\rho$ is given by $(\hat{O} \otimes \mathbb{I}) |\rho\rangle$, 
as described in Eqs.~(\ref{m1})--(\ref{m5}). 

\subsection{B. Temporal dynamics using the decomposed evolution superoperator}
For a time-independent Lindbladian, the temporal evolution of the system is then given by the equation, 
\begin{equation}
|\rho(t+\Delta t)\rangle = \exp({\tilde{\mathcal{L}}\Delta t})|\rho(t)\rangle = e^{\sum_k\tilde{\mathcal{L}}_k\Delta t}|\rho(t)\rangle = \mathcal{V}(\Delta t)|\rho(t)\rangle,
\end{equation}
where we have used the form of $\tilde{\mathcal{L}}$ from Eq.~(\ref{Lind2}).
It is important to note that the Lindbladian $\tilde{\mathcal{L}}$ is not Hermitian and all its eigenvectors do not necessarily correspond to physical states of the system. One that does is the null-space vector which corresponds to the steady-state of the system, i.e., $|\rho\rangle_{ss}$ which satisfies ${d|\rho\rangle_{ss}}/{dt} =\tilde{\mathcal{L}}|\rho\rangle_{ss}$ = 0. However, the open-system dynamics in the Lindblad picture ensures that the evolution superoperator $\mathcal{V}(t)$ maps any initial quantum state to a set of valid quantum states.


The second-order Suzuki-Trotter (ST) decomposition allows us to write the time-evolution superoperator, $\mathcal{V}(\Delta t)$, for small $\Delta t$, in terms of individual spin and cavity evolution superoperators, $\mathcal{V}_k(\Delta t)$, such that
\begin{eqnarray}
\mathcal{V}(\Delta t) &=& e^{\sum_k \tilde{\mathcal{L}}_k\Delta t} = e^{\tilde{\mathcal{L}}_N\frac{\Delta t}{2}} e^{\tilde{\mathcal{L}}_{N-1}\frac{\Delta t}{2}}\cdots e^{\tilde{\mathcal{L}}_{2}\frac{\Delta t}{2}} e^{\tilde{\mathcal{L}}_1 \Delta t}
e^{\tilde{\mathcal{L}}_{2}\frac{\Delta t}{2}} \cdots e^{\tilde{\mathcal{L}}_{N-1}\frac{\Delta t}{2}} e^{\tilde{\mathcal{L}}_N\frac{\Delta t}{2}} + \mathbb{O}(\Delta t^3),\nonumber\\
&\approx& \mathcal{V}_N({\Delta t}/{2})\mathcal{V}_{N-1}({\Delta t}/{2})\cdots\mathcal{V}_2({\Delta t}/{2})\mathcal{V}_1(\Delta t)\mathcal{V}_2({\Delta t}/{2})\cdots\mathcal{V}_{N-1}({\Delta t}/{2})\mathcal{V}_N({\Delta t}/{2}),
\label{ST}
\end{eqnarray}
with errors, $\mathcal{E}_\textrm{ST}$, of the order of $(\Delta t^3)$. Each $\mathcal{V}_k(\Delta t)$ can be numerically solved for the $k^{th}$ spin and the cavity.
Therefore, the above ST decomposition for the evolution superoperator, $\mathcal{V}(\Delta t)$, allows the implementation of the Lindbladian dynamics in a time-adaptive manner.  For time $\Delta t$, this is achieved by sequentially applying the superoperator $\mathcal{V}_k(\Delta t)$ on the $k^{th}$ spin and the cavity, starting from $k = N$, such that all the terms in Eq.~(\ref{ST}) have been addressed. The entire process involves a double sweep through the spins in the ensemble, starting from the $N^{th}$ spin to the $1^{st}$ spin (first sweep) and then back to the $N^{th}$ spin (second sweep), similar to the approach taken in a conventional TEBD or time-dependent DMRG method.
However, we note that $\mathcal{V}_k(\Delta t)$ is not a unitary time evolution operator, nor is it local, as is often the case in conventional TEBD or DMRG sweeps. %
We also state that there is no special convention to numbering the spins in the ensemble, since there is no innate geometry in the system. 

\subsection{C. Renormalization of the superoperator space}
The key step in a typical renormalization method is truncation of the reduced density matrix space \cite{Schollwoeck2011}. In our case, this translates to the renormalization of the reduced superoperator.
Let us consider that the spin-cavity state is decomposed into two blocks, say $L$ and $R$, such that $L$ contains $N_1$ spins, and $R$ contains the rest $N-N_1$ spins and the cavity. Without loss of generality, let us assume, $N_1 < N/2$. Hence, $|\rho\rangle$ = $\sum_{k_L,k_R} p_{k_L,k_R} |k_L\rangle |k_R\rangle$. Here $|k_L\rangle$ = $|k\rangle^{\otimes N_1}$ 
and $|k_R\rangle$ = $|k\rangle^{\otimes N-N_1}|k_c\rangle$, where $|k_i\rangle = |k\rangle, \forall~i$.
The Schmidt decomposition is given by
$|\rho\rangle = \sum_{\tilde{k}=1}^\mathcal{K} \lambda_{\tilde{k}}~ |\tilde{k}_L \rangle |\tilde{k}_R\rangle$, where $\{\lambda_{\tilde{k}}\}$ are the Schmidt coefficients in descending order. $|\tilde{k}_L\rangle$ and $|\tilde{k}_R\rangle$ are the Schmidt vectors, which are the eigenvectors of the reduced superoperators, $\mathcal{R}_{L}$ = $\textrm{Tr}_{R}(|\rho\rangle\langle\rho|)$ and $\mathcal{R}_{R}$ = $\textrm{Tr}_{L}(|\rho\rangle\langle\rho|)$, respectively. Here, $\textrm{Tr}_{R(L)}$ implies partial trace over $R(L)$ subsystem. For pure states, the Schmidt rank ($\mathcal{K}$) determines the entanglement between the bipartitions $L$ and $R$. However, since $|\rho\rangle$ is a vectorized density matrix, $\mathcal{K}$ in this case can be interpreted as a measure of total correlations \cite{Datta2007}. Theoretically, $1 \leq \mathcal{K} \leq {r}$, where $r$ = $\textrm{min}(\textrm{rank}(\mathcal{R}_{L}),\textrm{rank}(\mathcal{R}_{R}))$ = $d^{N_1}$. Here  
$d^{N_1}$ is the dimension of the block $L$, and is smaller than the dimension of $R$, i.e., $d^{N_1} < d^{N-N_1}$.
For large $N$, the rank $\mathcal{K}$ can thus grow exponentially with $N_1$. 
However, in most physical situations, especially in systems with central-body spin-cavity interactions, $\mathcal{K}$ is small or alternatively $\lambda_{\tilde{k}}$ decays rapidly with $\tilde{k}$ \cite{Zwolak2004}. 
Thus, a significantly smaller number, $\mathcal{D} \ll r$, can be used to effectively describe the system, i.e.,
$|\rho\rangle = \sum_{\tilde{k}=1}^\mathcal{D} \lambda_{\tilde{k}}~ |\tilde{k}_L \rangle |\tilde{k}_R\rangle$.
Subsequently, the transformation superoperator, $\mathcal{U}_{r\times\mathcal{D}}$, with the first $\mathcal{D}$  eigenvectors of $\mathcal{R}_{L}$ as columns, is used to renormalize all quantum objects in the $r\times r$ superoperator space in block $A$ to the truncated $\mathcal{D}\times \mathcal{D}$ space, i.e., $\forall~\{\hat{\mathcal{O}}_{r\times r},|\sigma\rangle_r\} \in L$, $\mathcal{U}^{-1}\hat{\mathcal{O}}_{r\times r}~\mathcal{U}\rightarrow \hat{\mathcal{O}'}_{\mathcal{D}\times \mathcal{D}}$ and $\mathcal{U}^{-1}|\sigma\rangle_r \rightarrow |\sigma'\rangle_\mathcal{D}$. Hence, in the truncated basis, 
\begin{eqnarray}
&|\rho'\rangle = (\mathcal{U}^{-1}\otimes \mathcal{I}_{k_R})|\rho\rangle = \sum_{\tilde{k}_L,k_R} \tilde{p}_{\tilde{k}_L,k_R} |\tilde{k}_L\rangle |k_R\rangle,~~\textrm{where}\\
&\tilde{p}_{\tilde{k}_L,k_R} = \sum_{k_L=1}^r \mathcal{U}^{-1}_{\tilde{k}_L,k_L}p_{k_L,k_R}.
\end{eqnarray}
$\mathcal{I}$ is the identity superoperator. We note that such a transformation, in general, is not exact and the corresponding error is given by the quantity, $\mathcal{E} = 1 - \textrm{Tr}[\rho]=1-\langle \tilde{t}|\rho'\rangle$, where $|\tilde{t}\rangle$ = $(\mathcal{U}'^{-1} \otimes \mathbb{I}_{d_R} \otimes \mathbb{I}_{d_c})|t\rangle$ (here $|t\rangle$ is the initial superket of the trace operator such that $\langle t|\rho\rangle$ = 1). As expected, a  finite error affects the normalization of the density matrix in the superoperator space, and thus the accuracy of the method, as is the case with all tensor-network methods. The validation and success of any renormalization or tensor-network method is thus dependent on the accrued value of the truncation error.
During the implementation of each time-adaptive renormalization step, the error is minimized by choosing an optimal $\mathcal{D}$ and the superket is always normalized, i.e., $|\rho(0)\rangle = \frac{1}{\langle \tilde{t}|\rho(0)\rangle}|\rho(0)\rangle$. For exact transformations, where $\mathcal{D}$ = $\mathcal{K}$, the error $\mathcal{E} = 0$. The error increases as the total correlations in the spin ensemble increase. This includes both classical and quantum correlations, in addition to possible entanglement. Together with the superoperator space and the decomposed superoperators, the Schmidt decomposition and renormalization provides the basic theoretical tools to solve the dynamics as well as the steady states of the  Lindbladian. 

\subsection{D. Time-adaptive variational renormalization group method for open-system dynamics}

To implement the temporal evolution using the time-adpative variational renormalization group method, we start by constructing an initial system in the renormalized superoperator space. Let, us consider the initial superket, $|\rho(0)\rangle$, to be a product of individual spins and cavity. This is in tune with the fact that the physical state of the mesoscopic ensemble at the start may not contain any excitations or alternatively, can be in an incoherently pumped uncorrelated state. However, one may also begin with correlated initial states without affecting the variational renormalization method. Let us begin with an initial system, $|\rho(0)\rangle$, which consists of $2m$ $d$-level spins or emitters in the ensemble ($m < N/2$), inside a quantum cavity with $n_c$ Fock states. 
Let us label the spins such that $|\rho(0)\rangle$ consists of two blocks of spins, the first $m$ spins ($L$ block) represented in the basis, $|k_L\rangle$ = $|k_1\rangle\otimes|k_2\rangle\dots|k_{m}\rangle$, and the last $m$ spins ($R$ block) in the basis, $|k_{R}\rangle$ = $|k_{N-m+1}\rangle\dots|k_{N-1}\rangle\otimes|k_{N}\rangle$, along with the 
cavity state represented in the superket basis $|k_c\rangle$.
The total dimension of $|\rho(0)\rangle$ is $(d^{2m}n_c)^2$, where $d_L$ = $d^{2m}$, $d_R$ = $d^{2m}$, and $d_c$ = $n_c^2$ are the dimensions of the $L$-, $R$-, and the cavity blocks, respectively. The initial product superket in the above basis, with the cavity in the vacuum state, is given by,
\begin{equation}
|\rho(0)\rangle = \sum_{k_L,k_R,k_c} p_{k_L,k_R,k_c}
|k_L\rangle \otimes |k_{R}\rangle\otimes|k_c\rangle = 
|k_l\rangle_L \otimes |k_r\rangle_{R}\otimes|0\rangle_c, \textrm{where,}~~p_{k_l,k_l,0} = 1. 
\end{equation}
To sequentially construct the $N$-spin ensemble we add the next spin to the $L$ block, at the unoccupied $(m+1)^{th}$ site, such that 
$
|\rho(0)\rangle = 
|k_l\rangle_L\otimes |k'\rangle_{m+1} \otimes |k_r\rangle_{R}\otimes|0\rangle_c.
$
The $L$ block now consists of  $m+1$ spins. 
We consider the reduced superoperator of the $L$ block and the $(m+1)^{th}$ spin given by $\mathcal{R}_{L,m+1}$ = $\textrm{Tr}_{R,c}[|\rho(t)\rangle\langle\rho(t)|]$. 
The space of the reduced superoperator, $\mathcal{R}_{L,m+1}$, is then renormalized, i.e., only $\mathcal{D}$ 
eigenvectors, corresponding to $\mathcal{D}$ highest eigenvalues of $\mathcal{R}_{L,m+1}$, are used to contruct the transformation matrix $\mathcal{U}^{m+1\leftarrow m}_{\mathcal{D}\times d_Ld^2}$, which maps the $d_Ld^2$ dimensional space of $m+1$ spins to a truncated $\mathcal{D}~(\leq d_Ld^2)$ dimensional renormalized space. The reduced operator renormalization presented here is the same as the Schmidt decomposition mentioned earlier or in general to the singular value decomposition of the bipartite coefficient matrix, $p_{k_a,k_b}$, of any density matrix $|\rho\rangle$ = $\sum_{k_a,k_b}p_{k_a,k_b}|k_a\rangle|k_b\rangle$. 
We note that for $\mathcal{D} = \mathcal{K} \leq d_Ld^2$, the mapping is exact, where $\mathcal{K}$ is the Schmidt rank for the bipartite decomposition between the first $(m+1)$ spins and the rest. Also, for product superkets, as is the case of our initial system, only a single non-zero eigenvalue exists, and the map is exact even for $\mathcal{D}$ = 1. However, to elaborate the steps in the renormalization process, which will also be necessary in the temporal evolution, we stick to an arbitrary $\mathcal{D}$. We note that this transformation occurs at the step where $m$ spins in the $L$ block expands to $m+1$ (left arrow in the superscript of $\mathcal{U}$). Therefore, the renormalization can be written as,
\begin{eqnarray}
&|\rho(0)\rangle \rightarrow (\mathcal{U}'^{-1} \otimes \mathbb{I}_{d_R} \otimes \mathbb{I}_{d_c})|\rho(0)\rangle = (\mathcal{U}'^{-1} \otimes \mathbb{I}_{d_R} \otimes \mathbb{I}_{d_c})
|k_l\rangle_L\otimes |k'\rangle_{m+1} \otimes |k_r\rangle_{R}\otimes|0\rangle_c, \textrm{where,}~~ \mathcal{U}' = \mathcal{U}^{m+1\leftarrow m}_{\mathcal{D}\times d_Ld^2}&\nonumber\\
&|\rho(0)\rangle = \sum_{\tilde{k}_L=1}^{{d_L}} p_{\tilde{k}_L}|\tilde{k}_L\rangle \otimes |k_r\rangle_{R}\otimes|0\rangle_c,~ \textrm{where,}~~ d_L = \mathcal{D}.&
\end{eqnarray}
In the next step, one adds the $(m+2)^{th}$ spin to the $L$ block, and derives the superoperator 
$\mathcal{R}_{L,m+2}$ and the transformation matrix, $\mathcal{U}^{m+2\leftarrow m+1}_{\mathcal{D}\times d_Ld^2}$, and the space is again truncated to $\mathcal{D}$.
In a similar manner one can add the remaining spins in the ensemble, with the last transformation being $\mathcal{U}^{N-m\leftarrow N-m-1}_{\mathcal{D}\times d_Ld^2}$, with the final superket being, $|\rho(0)\rangle = |\tilde{k}\rangle_L \otimes |k\rangle_{R}\otimes|k\rangle_c$, where there are now $N-m$ spins in the renormalized $L$ block, and $m$ spins in the un-renormalized $R$ block. We note that $m$, which can ideally be chosen such that $\mathcal{D} = d^{2m}$, is the defining value in the numerical process. However, lower values of $m$ can also be chosen, as necessary, to improve the runtime of the code without strongly hurting the accuracy, as defined by $\mathcal{E}$.

We note that the reduced superoperator, $\mathcal{R}$, for a given $|\rho\rangle$, does not itself describe a quantum state, but rather provides an alternate set of basis superkets, arising from the singular value decomposition of the coefficient matrix, which describes the reduced space of the quantum superket $|\rho\rangle$. This allows one to approximately map all the relevant quantum superoperators, superkets, and operators to the renormalized subspace, and thus observables of the system are preserved, albeit accommodating for the error arising from the truncation. Another important point is that for more than one party, the description of any $|\rho\rangle$, in the product superket basis, does not directly correspond to the vectorized form of the density matrix, i.e.,  $|\rho\rangle_{A:B} = \sum_k p_k |k\rangle_{A:B} \neq \sum_{k_A,k_B} p_{k} |k_{A}\rangle|k_B\rangle$, where $k = k_A k_B$. This is due to the fact that $|i,j\rangle\langle i',j'| \rightarrow |iji'j'\rangle \neq |ii'jj'\rangle$. This is usually pertinent while solving $\mathcal{V}_k$, which is represented in the joint superket basis of the spin and cavity rather than the product of individual superket basis.
Hence, $\mathcal{V}_k$'s must be transformed to the product superket basis before implementation.

Once the initial renormalized superket, $|\rho(0)\rangle$, has been constructed, the time evolution of the system can be performed. This is achieved through two sweeps (left and right) as described in the following.

\subsection*{1. Left sweep~ $\xleftarrow{j=N\cdots 1}$}

Once the initial renormalized superket, $|\rho(0)\rangle$, has been constructed the time evolution superoperator, $\mathcal{V}(\Delta t)$ can be applied to evolve the superket by time $\Delta t$. Let us consider, without loss of generality, $m$ = 1, such that there are $N-1$ spins in the renormalized $L$ block, and the single $N^{th}$ spin in the $R$ block. All spins are unexcited and the initial cavity has no photons, or is in the vacuum state. We set $\mathcal{D}'$ as the desired truncation rank of the reduced superoperator or the truncated dimension of the renormalized superoperator space. The initial superket is then given by, $|\rho(0)\rangle = \sum_{\tilde{k}_L=1}^{{d_L}} p_{\tilde{k}_L}|\tilde{k}_L\rangle \otimes |k_r\rangle_{R=N}\otimes|0\rangle_c$. We apply the evolution operator that acts on the $N^{th}$ spin and the cavity, $\mathcal{V}_N(\Delta t/2)$, such that
\begin{equation}
|\rho(\Delta t)\rangle^{\xleftarrow{j=N}} = (\mathbb{I}_{d_L} \otimes \mathcal{V}_N(\Delta t/2)) |\rho(0)\rangle
=\sum_{\tilde{k}_L,k_R,k_c} p^{\xleftarrow{j=N}}_{\tilde{k}_L,k_R,k_c}
|\tilde{k}_L\rangle \otimes |k_{R}\rangle\otimes|k_c\rangle.
\label{tN}
\end{equation}
The left arrow in the superscript of the superket and coefficients indicate the ``left sweep'' (from $N^{th}$ to $1^{st}$ spin), and the number, $j$, above the arrow indicates that all the higher evolution superoperators, i.e., $\mathcal{V}_k(\Delta t/2), \forall~ k \geq j$, have already been applied. The superket after such a partial evolution is no longer a product between the spin ensemble and the cavity. 

Before, we can apply the next superoperator, $\mathcal{V}_{N-1}(\Delta t/2)$, on the $(N-1)^{th}$ spin, we need to first release it from the renormalized block $L$. This is achieved by the applying the inverse of the transformation done during a previous renormalization step to form the block, i.e., inverse map of $\mathcal{U}'$ = $\mathcal{U}^{N-1\leftarrow N-2}_{\mathcal{D}\times d_Ld^2}$. This gives us,
\begin{equation}
|\rho(\Delta t)\rangle^{\xleftarrow{N}} \Longrightarrow (\mathcal{U}'\otimes \mathbb{I}_{d_R} \otimes \mathbb{I}_{d_c})  |\rho(\Delta t)\rangle^{\xleftarrow{N}} = \sum_{\tilde{k}_L,k_{j},k_R,k_c} p'^{\xleftarrow{N}}_{\tilde{k}_L,k_{j},k_R,k_c}
|\tilde{k}_L\rangle\otimes |k_{j}\rangle \otimes |k_{R}\rangle\otimes|k_c\rangle, \textrm{where,}~~j = N-1.
\label{revtrans}
\end{equation}
The transformation is given by, $p'^{\xleftarrow{N}}_{\tilde{k}_L,k_{j},k_R,k_c}$ = $ \sum_{\tilde{k}_L}\mathcal{U}'_{\tilde{k}_L k_j,\tilde{k}_L} p^{\xleftarrow{N}}_{\tilde{k}_L,k_R,k_c}$, and subsequently releases the $(N-1)^{th}$ spin. In the numerical formulation of the problem, we note that the evolution superoperator acts locally on the two-body system consisting of the $(N-1)^{th}$ spin and the cavity, although the sites are nonlocal in Eq.~(\ref{revtrans}). This can be overcome in an efficient way, using a swap operation \cite{Stoudenmire2010} such that,
$\mathcal{S}_{j:R}|\rho(\Delta t)\rangle^{\xleftarrow{N}}: = \sum_{\tilde{k}_L,k_{j},k_R,k_c} p'^{\xleftarrow{N}}_{\tilde{k}_L,k_R,k_{j},k_c}
|\tilde{k}_L\rangle\otimes |k_{R}\rangle\otimes |k_{j}\rangle \otimes|k_c\rangle$. Alternatively, other approaches could be applied especially if one uses matrix product operators to define the open dynamics \cite{Schollwoeck2011,Orus2014}. Thereafter, $\mathcal{V}_{N-1}(\Delta t/2)$, can be applied locally to the $(N-1)^{th}$ spin and the cavity, such that
\begin{equation}
|\rho(\Delta t)\rangle^{\xleftarrow{N-1}} = (\mathbb{I}_{d_L} \otimes \mathbb{I}_{d_R} \otimes\mathcal{V}_{N-1}(\Delta t/2)) |\rho(\Delta t)\rangle^{\xleftarrow{N}}
=\sum_{\tilde{k}_L,k_R,k_j,k_c} p^{\xleftarrow{N-1}}_{\tilde{k}_L,k_R,k_j,k_c}
|\tilde{k}_L\rangle \otimes |k_{R}\rangle\otimes|k_{j}\rangle \otimes|k_c\rangle, \textrm{where,}~~j = N-1.
\label{timestep}
\end{equation}
Now one arrives at the key step in the time-adaptive variational renormalization process. Before, the $(N-2)^{th}$ spin is released from the $L$ block, the $(N-1)^{th}$ spin, must be renormalized to the $R$ block. This allows the superket, $|\rho(\Delta t)\rangle$, at any time step, to consist of no more than one un-renormalized (free) spin and the cavity. The rest of the spins are renormalized to either the $L$ or $R$ blocks, similar to a standard sweep in TEBD and t-DMRG, apart from the fact that here the free objects are the density matrix of a nonlocal, hybrid spin-boson pair instead of a local pair of fermions. To renormalize the $(N-1)^{th}$ spin, we obtain the reduced superoperator of the $R$ block and the $(N-1)^{th}$ spin, which is given by 
$\mathcal{R}_{R,j}$ = $\textrm{Tr}_{L,c}[|\rho(\Delta t)\rangle\langle\rho(\Delta t)|^{\xleftarrow{j}}]$, where $j$ = $N-1$. Again, the space of the superoperator $\mathcal{R}_{R,j}$, is renormalized by keeping the $\mathcal{D}$ = $\min(\mathcal{D}',\textrm{dim}(\mathcal{R}_{R,j}))$ eigenvectors corresponding to the $\mathcal{D}$ highest eigenvalues. 
The choice of $\mathcal{D}$ here ensures that truncation only sets in when the rank of $\mathcal{R}_{R,j}$ exceeds a preset value $\mathcal{D}'$. Hence, truncation is only performed at higher ranks of the reduced superoperator.
Subsequently, we obtain the transformation matrix, $\mathcal{U}^{j\rightarrow j+1}_{\mathcal{D}\times d_Ld^2}$, such that upon renormalizing the superket in Eq.~(\ref{timestep}), we obtain
\begin{equation}
|\rho(\Delta t)\rangle^{\xleftarrow{N-1}} \Longrightarrow (\mathbb{I}_{d_L} \otimes    \mathcal{U}'^{-1}\otimes \mathbb{I}_{d_c})  |\rho(\Delta t)\rangle^{\xleftarrow{N-1}} = \sum_{\tilde{k}_L,\tilde{k}_R,k_c} p'^{\xleftarrow{N-1}}_{\tilde{k}_L,\tilde{k}_R,k_c}
|\tilde{k}_L\rangle\otimes |\tilde{k}_{R}\rangle\otimes|k_c\rangle, \textrm{where,}~~\mathcal{U}'=\mathcal{U}^{j\rightarrow j+1}_{\mathcal{D}\times d_Ld^2}.
\label{fortrans}
\end{equation}
This completes the set of steps involved in applying the decomposed time-evolution superoperator, $\mathcal{V}_{N-1}(\Delta t/2)$, on the $(N-1)^{th}$ spin, and is thus repeated for each of the $j$ spins, $1 < j < N-1$, as a part of the first sweep. The steps can be reiterated for all spins during the first sweep as summarized in the following, starting from the superket, $|\rho(\Delta t)\rangle^{\xleftarrow{j+1}}$ = $\sum_{\tilde{k}_L,\tilde{k}_R,k_c} p'^{\xleftarrow{j+1}}_{\tilde{k}_L,\tilde{k}_R,k_c}
|\tilde{k}_L\rangle\otimes |\tilde{k}_{R}\rangle\otimes|k_c\rangle$, where $j = N-2$ gives us the superket in Eq.~(\ref{fortrans}).
\begin{eqnarray}
&1)~ \textrm{Release the j}^{th}~ \textrm{spin from}~L~\textrm{block:}\nonumber\\
&(\mathcal{U}'\otimes \mathbb{I}_{d_R} \otimes \mathbb{I}_{d_c})  |\rho(\Delta t)\rangle^{\xleftarrow{j+1}} = \sum_{\tilde{k}_L,k_{j},\tilde{k}_R,k_c} p''^{\xleftarrow{j+1}}_{\tilde{k}_L,k_{j},\tilde{k}_R,k_c}
|\tilde{k}_L\rangle\otimes |k_{j}\rangle \otimes |\tilde{k}_R\rangle\otimes|k_c\rangle, \textrm{where,}~~\mathcal{U}'=\mathcal{U}^{j\leftarrow j-1}_{\mathcal{D}\times d_Ld^2},\label{s1}\\\nonumber\\
&2)~ \textrm{Evolve the j}^{th}~ \textrm{spin, after swap, using the superoperator}~\mathcal{V}_j(\Delta t/2)\nonumber\\
&|\rho(\Delta t)\rangle^{\xleftarrow{j}} = (\mathbb{I}_{d_L} \otimes \mathbb{I}_{d_R} \otimes\mathcal{V}_{N-1}(\Delta t/2)) |\rho(\Delta t)\rangle^{\xleftarrow{j+1}}
=\sum_{\tilde{k}_L,\tilde{k}_R,k_j,k_c} p^{\xleftarrow{j}}_{\tilde{k}_L,\tilde{k}_R,k_j,k_c}
|\tilde{k}_L\rangle \otimes |\tilde{k}_R\rangle\otimes|k_{j}\rangle \otimes|k_c\rangle,\label{s2}\\\nonumber\\
&3)~ \textrm{Renormalizing the j}^{th}~ \textrm{spin to}~R~\textrm{block:}\nonumber\\
&(\mathbb{I}_{d_L} \otimes    \mathcal{U}'^{-1}\otimes \mathbb{I}_{d_c})  |\rho(\Delta t)\rangle^{\xleftarrow{j}} = \sum_{\tilde{k}_L,\tilde{k}_R,k_c} p'^{\xleftarrow{j}}_{\tilde{k}_L,\tilde{k}_R,k_c}
|\tilde{k}_L\rangle\otimes |\tilde{k}_{R}\rangle\otimes|k_c\rangle, \textrm{where,}~~\mathcal{U}'=\mathcal{U}^{j\rightarrow j+1}_{\mathcal{D}\times d_Ld^2}.\label{s3}\\\nonumber
\end{eqnarray}

Following the above steps, $\mathcal{V}_j(\Delta t/2)$, can be applied to all the spins in the first sweep, till $j = 1$, where all but the $1^{st}$ spin is in the $R$ block. At this point, following a swap $\mathcal{S}_{1:R}$, the time evolution is given by,
\begin{equation}
|\rho(\Delta t)\rangle^{\xleftarrow{j=1}} = (\mathbb{I}_{d_R} \otimes\mathcal{V}_1(\Delta t)) |\rho(0)\rangle
=\sum_{\tilde{k}_L,k_R,k_c} p^{\xleftarrow{j=1}}_{\tilde{k}_R,k_j,k_c}
|\tilde{k}_R\rangle \otimes |k_{j}\rangle\otimes|k_c\rangle.
\end{equation}

\subsection*{2. Right sweep~ $\xrightarrow{j=1\cdots N}$}

This sets the stage for the second sweep, the ``right sweep'' (from $N^{th}$ to $1^{st}$ spin), where the $j^{th}$ spin is now released from the $R$ block, evolved using the superoperator $\mathcal{V}_j(\Delta t/2)$, and then renormalized to the $L$ block, for $1<j < N$. The right arrow in the superscript of the superket and coefficients indicate the right sweep, and the number, $j$, above the arrow indicates that all the lower evolution superoperators, i.e., $\mathcal{V}_k(\Delta t/2), \forall~ k \leq j$, have already been applied.
This is the reverse set of operations of the steps outlined in Eqs.~(\ref{s1})--(\ref{s3}). 

\begin{eqnarray}
&1)~ \textrm{Release the j}^{th}~ \textrm{spin from}~R~\textrm{block:}\nonumber\\
&(\mathbb{I}_{d_L} \otimes\mathcal{U}'\otimes \mathbb{I}_{d_c})  |\rho(\Delta t)\rangle^{\xrightarrow{j-1}} = \sum_{\tilde{k}_L,k_{j},\tilde{k}_R,k_c} p''^{\xrightarrow{j-1}}_{\tilde{k}_L,k_{j},\tilde{k}_R,k_c}
|\tilde{k}_L\rangle\otimes |k_{j}\rangle \otimes |\tilde{k}_R\rangle\otimes|k_c\rangle, \textrm{where,}~~\mathcal{U}'=\mathcal{U}^{j\rightarrow j+1}_{\mathcal{D}\times d_Ld^2},\label{t1}\\\nonumber
\end{eqnarray}
We note that here $\mathcal{U}^{j\rightarrow j+1}_{\mathcal{D}\times d_Ld^2}$ is the transformation superoperator obtained during the left sweep in Eq.~(\ref{s3}), for the operation leading to the $j^{th}$ spin being renormalized to the $R$ block.

\begin{eqnarray}
&2)~ \textrm{Evolve the j}^{th}~ \textrm{spin, after swap, using the superoperator}~\mathcal{V}_j(\Delta t/2)\nonumber\\
&|\rho(\Delta t)\rangle^{\xrightarrow{j}} = (\mathbb{I}_{d_L} \otimes \mathbb{I}_{d_R} \otimes\mathcal{V}_{N-1}(\Delta t/2)) |\rho(\Delta t)\rangle^{\xrightarrow{j-1}}
=\sum_{\tilde{k}_L,\tilde{k}_R,k_j,k_c} p^{\xrightarrow{j}}_{\tilde{k}_L,\tilde{k}_R,k_j,k_c}
|\tilde{k}_L\rangle \otimes |\tilde{k}_R\rangle\otimes|k_{j}\rangle \otimes|k_c\rangle,\label{t2}
\end{eqnarray}
\begin{eqnarray}
&3)~ \textrm{Renormalizing the j}^{th}~ \textrm{spin to}~L~\textrm{block:}\nonumber\\
&( \mathcal{U}'^{-1}\otimes\mathbb{I}_{d_R} \otimes    \mathbb{I}_{d_c})  |\rho(\Delta t)\rangle^{\xrightarrow{j}} = \sum_{\tilde{k}_L,\tilde{k}_R,k_c} p'^{\xrightarrow{j}}_{\tilde{k}_L,\tilde{k}_R,k_c}
|\tilde{k}_L\rangle\otimes |\tilde{k}_{R}\rangle\otimes|k_c\rangle, \textrm{where,}~~\mathcal{U}'=\mathcal{U}^{j\leftarrow j-1}_{\mathcal{D}\times d_Ld^2}.\label{t3}\\\nonumber
\end{eqnarray}

For $j = N$, the time evolution operation given by Eq.~(\ref{tN}) is performed. At the end of the double sweep, the spin-cavity ensemble is evolved in time $\Delta t$, and thus forms the initial superket for further evolution. 

\section{II. Spin and cavity observables in the renormalized superoperator space}

One can use the time-adaptive process to calculate local observables of the mesoscopic spin ensemble cavity system. Let us begin with the observables related to the quantum cavity, $\hat{O}_c$. The expectation value of the observable, $\hat{O}_c$, is given by $\langle\hat{O}_c\rangle$ = $\textrm{Tr}[\hat{O}_c \rho(t)]$. In the superoperator space, $\hat{O}_c \rho(t) \rightarrow (\hat{O}_c \otimes \mathbb{I}_{d_c}) |\rho(t)\rangle$ = $\hat{\mathcal{O}}_c |\rho(t)\rangle$, and therefore, $\langle\hat{O}_c\rangle$ = $\langle t|\hat{\mathcal{O}}_c |\rho(t)\rangle$, where $|t\rangle$ is the trace superket. At the end of each double sweep, say after time $t$ = $n\Delta t$, the superket is given by $|\rho(t)\rangle = 
\sum_{\tilde{k}_L,k_j,k_c} p_{\tilde{k}_L,k_j,k_c}
|\tilde{k}_L\rangle \otimes |k_{j}\rangle\otimes|k_c\rangle$, where $j = N$. Hence, the expectation value $\langle\hat{O}_c\rangle$ can be expressed as,
\begin{equation}
|\tilde{\rho}(t)\rangle = (\mathbb{I}_{d_s}\otimes\hat{\mathcal{O}}_c)\sum_{\tilde{k}_s,k_c} p_{\tilde{k}_s,k_c}
|\tilde{k}_s\rangle\otimes|k_c\rangle,
\end{equation} 
where $|\tilde{k}_s\rangle$ = $|\tilde{k}_L\rangle|{k}_N\rangle$, and $\tilde{k}_s = \tilde{k}_L k_N$. The observable is then given by, $\langle\hat{O}_c\rangle$ = $\langle \tilde{t}|\tilde{\rho}(t)\rangle$, where $\langle \tilde{t}|$ =
$\langle \tilde{t}_s|\otimes\langle t_c|$ is the renormalized trace superket. 

Calculating the local observable for the spins in the ensemble follows a similar notion but is a little more involved in terms of implementation. After time $t$, a separate sweep through the spins in the ensemble is run, similar to the first sweep in Eqs.~(\ref{s1})--(\ref{s3}), but now without the time-evolution. For the $j^{th}$ spin, the superket is of the form, $|\rho(t)\rangle$ =
$\sum_{\tilde{k}_L,k_{j},\tilde{k}_R,k_c} p''^{\xleftarrow{j+1}}_{\tilde{k}_L,k_{j},\tilde{k}_R,k_c}
|\tilde{k}_L\rangle\otimes |k_{j}\rangle \otimes |\tilde{k}_R\rangle\otimes|k_c\rangle$, after releasing the $j$ spin in Eq.~(\ref{s1}). For a local observable, $\hat{S}_j$, in the superoperator form, $\hat{\mathcal{S}}_j$ = $\hat{S}_j\otimes \mathbb{I}_{d_j}$, the expectation value is given by,
\begin{eqnarray}
&|\rho'(t)\rangle_{\langle\hat{S}_j\rangle} = (\mathbb{I}_{d_L}\otimes\hat{\mathcal{S}}_j\otimes\mathbb{I}_{d_R}\otimes\mathbb{I}_{d_c})\sum_{\tilde{k}_L,k_{j},\tilde{k}_R,k_c} p''^{\xleftarrow{j+1}}_{\tilde{k}_L,k_{j},\tilde{k}_R,k_c}
|\tilde{k}_L\rangle\otimes |k_{j}\rangle \otimes |\tilde{k}_R\rangle\otimes|k_c\rangle\\\nonumber\\
&\langle\hat{S}_j\rangle = \langle \tilde{t}|\rho'(t)\rangle_{\langle\hat{S}_j\rangle},~\textrm{where}~~\langle \tilde{t}| =
\langle \tilde{t}_L|\otimes\langle t_j|\otimes\langle \tilde{t}_R|\otimes\langle t_c|~~\textrm{is the renormalized trace superket}.
\end{eqnarray}
By sweeping through different $j$ spins the local spin observables can be easily calculated. We note that no renormalization is necessary during this sweep as the transformation superoperators obtained during the previous time-evolution are used. As such, the sweep to calculate the observables are much faster to implement. 

\section{III. Higher order correlation functions}

In the main text, we consider two key observables for the cavity: the average cavity photon number at time $t$, given by $\langle \hat{a}^{\dag}_c\hat{a}_c\rangle$ and the equal-time second order correlation function,  $g_2(t)$ = $\langle \hat{a}^{\dag 2}_c(t) \hat{a}_c^2(t)\rangle$/$\langle \hat{a}^{\dag}_c(t)\hat{a}_c(t)\rangle^2$. To measure the average excitation in the spins we calculate,  $\langle \sigma^+_i\sigma^-_i\rangle_{\omega_i = \omega_c}$, where spins in the subensemble with resonant transitions are considered. 
\begin{figure}[b]
\includegraphics[width=5in,angle=00]{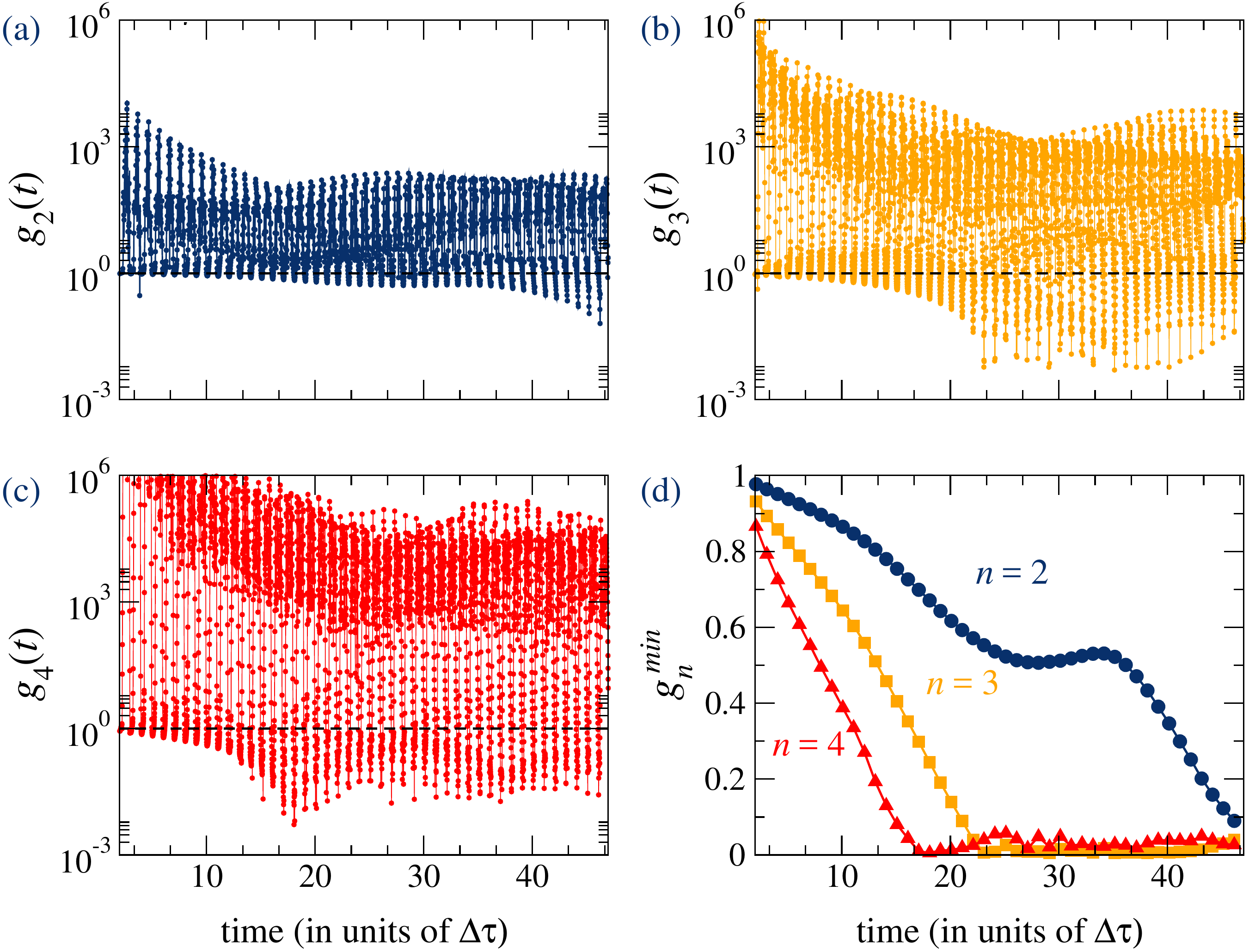}
\caption{(a)-(c) Equal-time higher order correlation functions, $g_n(t)$, for the nonclassical periodic pulse train obtained in the main text, for $n$ = 2, 3, 4, for a mesoscopic ensemble containing $N$ = 105 spins. (d) The minimum of $g_n(t)$ corresponding to each pulse in the train. As can be seen, the higher order correlation functions for $n$ = 2, 3, 4 all drop considerably below 1.}
\label{sup_fig1}
\end{figure}
While $\langle \hat{a}^{\dag}_c\hat{a}_c\rangle$ and $\langle \sigma^+_i\sigma^-_i\rangle$ give us the necessary information about the excitations stored in the cavity and the ensemble, respectively, the nonclassicality of the cavity field and the underlying photon statistics is given by $g_2(t)$. It is known that for coherent light the photon statistics follows a Poissonian distribution and $g_2(t)$ = 1. For thermal sources, the light is super-Poissonian and $g_2(t)$ = 2, and in general for all classical sources, $g_2(t) \geq $ 1.
However, for sub-Poissonian photon statistics, i.e.,  $g_2(t) < 1$, there is no classical analogue, and the photon emission essentially exhibits signatures of the quantum nature of light. This is the key to obtaining quantum features such as antibunching and single-photon sources. Over the years, the photon statistics of light has been one of the key parameters in studying nonclassicality of the electromagnetic field in quantum optics and cavity QED. 

In the context of designing nonclassical and single-photon sources, higher order correlation functions have shifted into the focus of attention. This is in response to phenomena such as unconventional photon blockade, where higher order photon processes may exist even in the absence of two photon processes. The equal time higher order correlations functions, at time $t$, are defined as 
$g_n(t)$ = $\langle \hat{a}^{\dag n}_c(t) \hat{a}_c^n(t)\rangle$/$\langle \hat{a}^{\dag}_c(t)\hat{a}_c(t)\rangle^n$. In instances of unconventional photon blockade, while $g_2(t) < 1$, implying antibunching and low probability for two photon emission, $g_3(t) > 1$, allowing for bunching in three photon emission \cite{Radulaski2017}. 
For an efficient single photon source, all higher order correlation functions must therefore satisfy the relation, $g_n(t) < 1$, thus allowing for a higher probability of single photon emission \cite{Carreno2016}.

For the nonclassical photon pulse obtained in our model, as shown in the main text, we observe that all higher order correlation functions are less than unity. In Fig.~\ref{sup_fig1} in the supplemental material, we demonstrate $g_n(t)$ and its minimum corresponding to each pulse, up to $n$ = 4, for an ensemble containing $N = 105$ spins. Here, all higher order photon emissions are antibunched with signifcantly lower probabilities as compared to single photon emission.

\section{IV. Sources of errors and implementation of the time-adaptive renormalization method}

The primary source of numerical errors, as with all renormalization group methods, arises from the Suzuki-Trotter decomposition of the evolution superoperator ($\mathcal{E}_\textrm{ST}$) and the truncation error during the renormalization of the reduced superoperator space ($\mathcal{E}$). 
We note that $\mathcal{E}_\textrm{ST}$ is not negligible in our model, as is often the case in one-dimensional quantum spin systems with short-range interactions, where the decomposition can be broken into odd and even sites that commute with each other. Nonetheless, the time-step ($\Delta t$) can be chosen to be significantly below the intrinsic time-scale of the spin-cavity dynamics, as governed by the spin and cavity frequencies, coupling strength and loss terms. 
The main aim is to set $\mathcal{E}_\textrm{ST}$ a few orders of magnitude below the truncation error $\mathcal{E}$. In our problem, this was done by comparison with exact dynamics for small $N$, since the $\mathcal{E}_\textrm{ST}$ can at most scale linearly with $N$. Excellent agreement was already obtained for $\Delta t$ in the order of $10^{-2}$ times the characteristic time scale, $2\pi/\Delta\omega$. 

For the problem on hand, we find that the main source of error in the method is the 
truncation error which increases as the total correlations in the superoperator space increase. 
In our case, the relevant correlations correspond to those generated among the spins in the ensemble, as the cavity space is not renormalized. However, we note that the spins are initially uncorrelated, and the correlations build up with time due to the spin-cavity interaction. Moreover, these correlations are not related to only the entanglement in the system. As the dynamics is open and the evolving system is mixed, both classical and quantum correlations are important. These correlations are higher for more excitations in the spins of the ensemble. For an ensemble of unexcited spins, the collective spin polarization vector rests at the south pole of an abstract Bloch sphere, i.e., $J_z = -1$, where $J_z$ is the collective $\sigma_z/2$ operator. As the spins are excited due to the cavity driving, the polarization vector moves towards the equator of the sphere, and the correlations in the ensemble increase. To observe nonclassical effects, it is important not to drive the cavity too strongly, such that it bosonizes the ensemble and cavity field and we enter the semiclassical regime. This amounts to the upper branch of the amplitude bistability curve in any hybrid spin-cavity systems.
In our work, the system is driven to move the polarization vector away from its initial state and beyond the regime of Holstein-Primakoff approximation ($J_z \approx -1$). However, the driving is not too strong such as to prevent large correlations to build up, which go beyond the capabilities  of any renormalization method. In our work, the driving is confined to excitations of the order of $J_z \lesssim -0.85$ such that the truncation error $\mathcal{E}$ can be set on the order of $10^{-9}$, with a choice of $\mathcal{D} \lesssim 200$. In principle, $J_z \approx -0.5$, can be achieved with reasonable effort, although such high driving is not relevant for the present study.  For smaller ensembles or weaker driving strengths, the error can be much smaller even for significantly lower value of $\mathcal{D}$. Another source of error in the numerical implementation of the method, specific to spin-cavity systems, is the finite dimension $n_c$ of the Fock basis describing the cavity. While in principle the cavity is infinite-dimensional, a fixed number of Fock states is chosen in tune with the driving. In our work, $n_c$ is chosen such that the probability of occupation in the highest Fock state, during the dynamics, is not above $10^{-9}$, i.e., Tr$[|n_c\rangle\langle n_c|\rho_c]\approx10^{-9}$, where $\rho_c$ is the reduced cavity density matrix. Depending on the driving strength, $n_c$ can be chosen to represent few ($n_c$ = 2--10) to several ($n_c >$ 100) Fock states, although the cavity field may start behaving classically as the photon number increases. Hence, in the interesting nonclassical regime, where the driving does not saturate the spin-cavity excitations and forces them towards the semiclassical limit, the total correlations in the spin ensemble are reasonably low and the variational renormalization performs well with sufficient accuracy.

The numerical time-adpative renormalization group method for mesoscopic spin-cavity systems was developed in C++, using the Armadillo 8.5 library for linear algera \cite{Sanderson2016}. The initial benchmark and accuracy of the method were established by comparing the results with exact Master equation solutions for smaller systems, developed in Python, using the QuTip library \cite{Johansson2012}. For systems with up to $N$ = 10 spins in the ensemble, and a small number of Fock states in the cavity, the full quantum solutions can be efficiently obtained using the master equation solver in QuTip. This serves as a good tool to benchmark the accuracy of our renormalization method. For this purpose, we considered the spin-cavity interactions in both the good cavity and bad cavity regime, for ensembles with inhomogeneous broadening, and  looked at both local and collective observables. Initial studies on the Rabi oscillations in the cavity field, spin magnetization and collective spin operators in inhomogeneously broadened systems showed that the time-adaptive renormalization group could reproduce the exact solutions with excellent accuracy using only a very small truncated dimension $\mathcal{D}$. For instance, for an inhomogeneous spin ensemble of 8 spins (initially unexcited) in the good-cavity regime, driven by a weak coherent drive, the average photon number ($\langle \hat{a}^{\dag}_c\hat{a}_c\rangle$) and the spin magnetization ($\langle S_z\rangle$) can be estimated with an accuracy of four significant decimal places by setting $\mathcal{D}$ to a very low value of 16, with the time-step being $1/100^{th}$ of the Rabi period. Moreover, in a stronger driving regime, where the driving strength $\eta > \kappa,\gamma$, where $\kappa$ and $\gamma$ are the cavity and spin loss terms, similar accuracy can be achieved by setting $\mathcal{D}\leq$ 64. 
The model could be readily extended to contain larger spin ensembles by linearly increasing $\mathcal{D}$ in our renormalization method, without any increase in the truncation error $\mathcal{E}$, and thus without any detrimental effect on the overall accuracy. In particular, for an homogeneous ensemble of $N$ = 200 spins in the strong coupling and driving regime, such that $\Omega \approx 10\kappa$, where $\Omega$ is the collective coupling strength and $\eta \approx 15\kappa$,  truncation errors $\mathcal{E} \leq 10^{-9}$ could be achieved with dimension as small as $\mathcal{D}$ = 100.
These results prove extremely promising for extending our method to larger spin ensembles.

We also considered a more complex case, i.e., the superradiance to subradiance transition with increasing inhomogeneous broadening of the spin ensemble in the bad cavity regime \cite{Temnov2005}.  The transition frequencies of the initially excited spin ensemble were chosen to follow a Gaussian distribution with a fixed half-width. 
Using exact quantum solutions for a small spin ensemble of $N$ = 6 spins we benchmarked our time-adpative renormalization method: 
For time steps $1/5^{th}$ of the characteristic time scale $2\pi/\kappa$, where $\kappa$ is the dominating term, and setting $\mathcal{D} \leq$  64, we found that the collective spin observables,  $J = \sum_{i,j}\langle \sigma^+_i \sigma^-_j\rangle$, can be estimated with an accuracy of four significant digits and two decimal places. The accuracy can be improved by considering smaller $\Delta t$ and higher $\mathcal{D}$. 
The transition from superradiant to subradiant states is well captured by these collective operators. For higher number of spins, the spin ensemble is highly correlated as the spins decay from the excited to the unexcited state in the absence of external driving, but still
the superradiant-subradiant transition in the collective spin state can be accurately captured using our method, albeit using more resources. 

\end{widetext}


\begin{thebibliography}{100}
\bibitem{comment}{The number of subensembles or distinct frequencies in the spectral comb, $m$, affects the sharpness of the pulse train of light. Fewer $m$ may lead to weaker rephasing around the central frequency and loss of sharpness in the revival. Increasing $m$ leads to sharper pulses, but subensembles too far detuned from the central frequency no longer contribute to the dynamics due to the very weak coupling with the cavity. In our study, we choose $m$ = 7 as the optimal number of frequencies that produce a sharp pulse train of photons.}
\end{thebibliography}
\end{document}